\documentclass[modern,twocolumn]{aastex62}

\usepackage{textcomp}
\usepackage{amsmath}

\received{}
\revised{}
\accepted{}

\submitjournal{Icarus}

\begin{document}

\title{A Challenge for Martian Lightning: Limits of Collisional Charging at Low Pressure}

\correspondingauthor{Gerhard Wurm}
\email{gerhard.wurm@uni-due.de}

\author{Gerhard Wurm}
\author{Lars Schmidt}
\author{Tobias Steinpilz}
\author{Lucia Boden}
\author{Jens Teiser}

\affiliation{University of Duisburg-Essen}

\begin{abstract}

Collisional charging is {one potential} initial step in generating lightning.
In this work, we study the charging of colliding monodisperse, spherical basalt grains depending on ambient pressure. We used grains of 1.0 to 1.2 mm in one set and 2.0 to 2.4 mm in another set. We varied the ambient pressure between 0.03 mbar and 80 mbar. This especially includes Martian pressure being 6 mbar on average. At a few mbar the net charge gathering on colliding grains has a minimum. A smooth incline in charging occurs for larger pressures. Toward lower pressure the charge increases steeply. The pressure dependence is in agreement to a model where the maximum charge is limited by a gas discharge occurring between two charged colliding grains shortly after or before a collision. The capability of building up charge is at a minimum exactly in the range of Martian pressures. The charges on grains are at least a factor 5 smaller than at the highest pressure tested and still smaller compared to ambient pressure on Earth. This implies that on Mars collisional charging and the potential of subsequent generation of lightning {or other large scale discharges} are strongly reduced compared to Earth. This might result in less frequent and less energetic lightning on Mars.

\end{abstract}

\keywords{Mars, charging, dust, lightning}

\section{Introduction}

Collisions of dust and sand sized particles are well known to lead to charging. Many mechanisms can be responsible depending on the materials, adsorbates and grain sizes involved \citep{lacks2011, Waitukaitis2014, Lee2018, Haeberle2018}.
Identical grains (spherical, same size, same material) can charge in collisions as well and highly \citep{Jungmann2018}. Induced dipoles have been proposed to promote charging in this case \citep{Siu2014, Yoshimatsu2016b, Yoshimatsu2017}.
Whatever the physical mechanism might be in detail, charging of individual grains in collisions on a small scale {is one important} first step in the generation of large scale discharges, e.g. by lightning {especially in dusty environments \citep{Harrison2016}}. 
{Certainly, a number of other mechanisms exist to electrify the atmosphere and particles within, ranging from ionization by cosmic rays and photoelectric charging over inductive charging to fracto-emission \citep{Yair2008, Bazilevskaya2008}. We concentrate on the role of collisional charging here. Even going beyond solar system planets,}
it has recently been shown in laboratory experiments that collisional charging also works at high temperatures relevant for extrasolar giant planet atmospheres \citep{Harper2018}. 

{In any case, after charging and charge separation, discharges in planetary atmospheres come in some variety \citep{Roussel2008}.} 
During a common thunderstorm on Earth, upon volcanic eruptions, or in dust storms, charge separation generates high voltages and produces lightning eventually \citep{Mason1988}. The generation of volcanic like lightning has also been shown in analog laboratory experiments to work well incorporating different grains sizes \citep{Cimarelli2014}. 

{Lightning requires a certain electrical field though to occur and in the conventional breakdown process the breakthrough voltage} depends strongly on the ambient pressure. The atmospheric pressure on Earth is between a few hundred mbar and 1 bar in the troposphere. It can be considerably different in extraterrestrial settings. 
A special case in the solar system is Mars. Here, the ambient pressure at the surface is low with 6 mbar on average. It varies from below 1 mbar at the highest volcanic elevations to beyond 10 mbar in deep valleys \citep{Wolkenberg2010, Lewis2003}. The question is if this low pressure is beneficial or disadvantageous for generating large scale discharges.

Generally, dust motion should also go with charging on Mars \citep{Harrison2016, Neakrase2016}. {\citet{Farrell1999a} simulated the capability of the Martian atmosphere to generate discharges concluding that it should be possible to some extent}
and \citet{Eden1973} already reported the generation of cm-sized discharges in a laboratory setting at low ambient pressure. At first glance, this comes as no surprise. According to standard discharge physics a certain (low) pressure range favours discharges due to the capability of generating electron and ion avalanches (see details in modeling section). And with this in mind, one might argue that Martian pressure should make large scale lightning much easier compared to Earth. More specific, \citet{Melnik1998} and \citet{Kok2009b} quantify the breakthrough voltage to be about a factor 100 lower on Mars compared to Earth or on the order of 20 to 25 kV/m compared to 3 MV/m. This should increase the chance for lightning.

However, Mars is not exactly known for frequent lightning observations. In fact, as far as even the mere existence of lightning or large scale discharges on Mars is concerned, no final statement can be given yet.  \citet{Ruf2009} were the first to announce ground based observations being consistent with the non-thermal microwave radiation originating from a Martian dust storm. However, \citet{Gurnett2010} analyzed 5 years of {\it Mars Express} data and could not find radio signatures of lightning. Also the 3 month search by \citet{Marin2012} did not show signals of lightning. Ground based observations attempting to confirm lightning are ongoing \citep{Majid2018}. 
It is curious that it is not even settled if lightning, related to dust activities, occurs on Mars. We take this as suggestion {to support} the idea that the low pressure might in fact reduce the likelihood for lightning rather than enhance it and study the pressure dependence of charging of individual grains.

We especially consider the primary charging of (two) colliding grains here which will be different comparing Earth and Martian conditions. We do not attempt to simulate all aspects of collisional charging, e.g. the low water content of the Martian atmosphere and the different composition which is mainly carbon dioxide on Mars, nor do we consider a potentially different set of minerals. 
We focus on the influence of the low atmospheric pressure on the maximum equilibrium charge on colliding grains. The idea is that a discharge by a breakthrough in the ambient atmosphere might already occur readily on a rather small scale between two colliding grains. If the grains get charged and separate from each other after a collision, discharge on a local scale might occur. This, in turn, prevents the generation of large scale voltage. 

It is not new that small scale breakdown in gaseous atmospheres might limit the maximum charge exchanged. \citet{Horn1992} were mostly reporting on the increased adhesion force going along with charging of two dissimilar materials but they also quantified the charges. They used silica on mica surfaces and study forces and charges while separating both surfaces after contact. Upon separations on the micrometer scale they observe partial discharge which they attribute to breakdown in the ambient atmosphere. Increasing the (nitrogen) pressure, discharge occurs at lower distances in agreement to breakdown theory (see below). Also in the context of adhesion, \citet{Broermann2012} observe the light associated with discharges upon detachment of micro-structured PDMS sheets in air. Along the same line of arguing upon gaseous discharge \citet{Matsuyama2018} recently discuss the maximum charge that a single particle up to the mm-size can gain in collisions with a metal target before a discharge limit is reached. Combining charging and discharging \citet{Haeberle2018} model the charge distributions observed in experiments of grains impacting different targets. Closely related to our work, \citet{Harper2016} consider a maximum charge related to discharge after particle collisions in experiments simulating triboelectric charging of volcanic ash.
 
Therefore, the possibility and occurence of discharge between two colliding grains separating from each other is not in question. We complement the existing work here by experiments and a simple model which show the charging capabilities limited by discharge under Martian conditions of low pressure. 

\section{Experiment}

\begin{figure}[h]
	\includegraphics[width=\columnwidth]{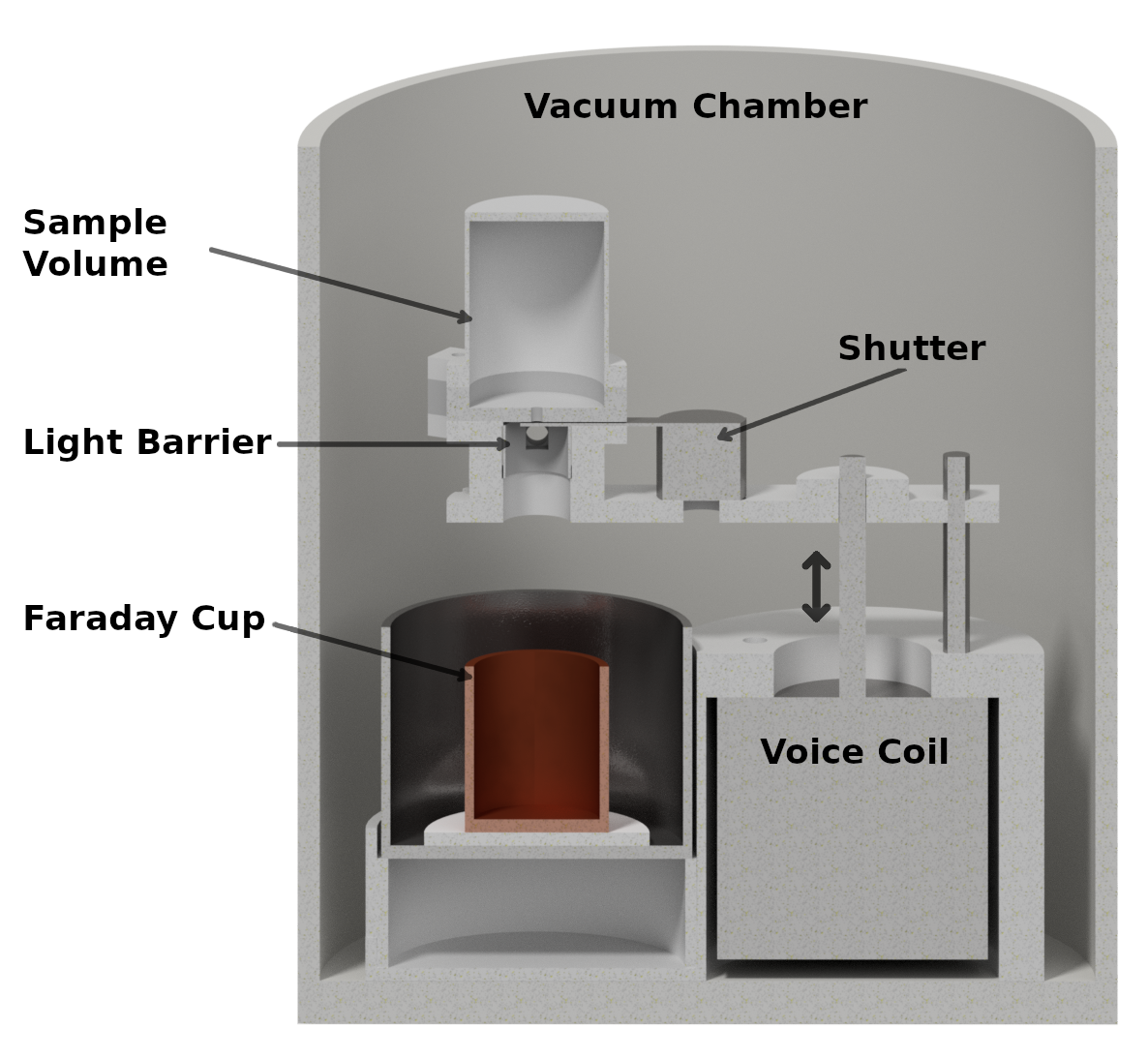}
	\caption{\label{fig.setup}Schematics of the experiment. A particle sample is vibrated by means of a voice coil. The walls of the sample volume are covered with the same particles. Individual grains can be extracted through a hole at the bottom. Their charge is measured by an electrometer.}
\end{figure}

A sketch of the experiment is shown in fig. \ref{fig.setup}.
Grains were vibrated for 30 minutes in a cylindrical container with 30 mm diameter and a height of 35 mm at 45 Hz {with an amplitude of about 1 mm.
These conditions were not chosen to match the dynamics of grain collisions under natural conditions, i.e. at Martian conditions. We only consider an equilibrium charge distribution and the influence of ambient pressure here. Different shaking will change the way grains acquire charge. With different charge patterns on the surface this might also influence the equilibrium charges but we consider this as secondary in the context of this work. Effects of the variations in the dynamics should be studied and are planned for the future}. The cylinder consisted of ABS but the inside was covered with a layer of the same particles that are vibrated. As part of a series to measure the collisional charging of identical grains, we took care to avoid glue facing the vibrated grains though this is of minor importance in the context of this paper. The cylinder had a conical bottom to allow grains to settle and move into the center.

As samples we used spherical basalt grains with a small size variation of 10 \% or between 1 to 1.2 mm in size and 
2 to 2.4 mm for a second sample.
{Bright field images of the grains can be seen in fig. \ref{fig.foto}. 
\begin{figure}[h]
	\includegraphics[width=\columnwidth]{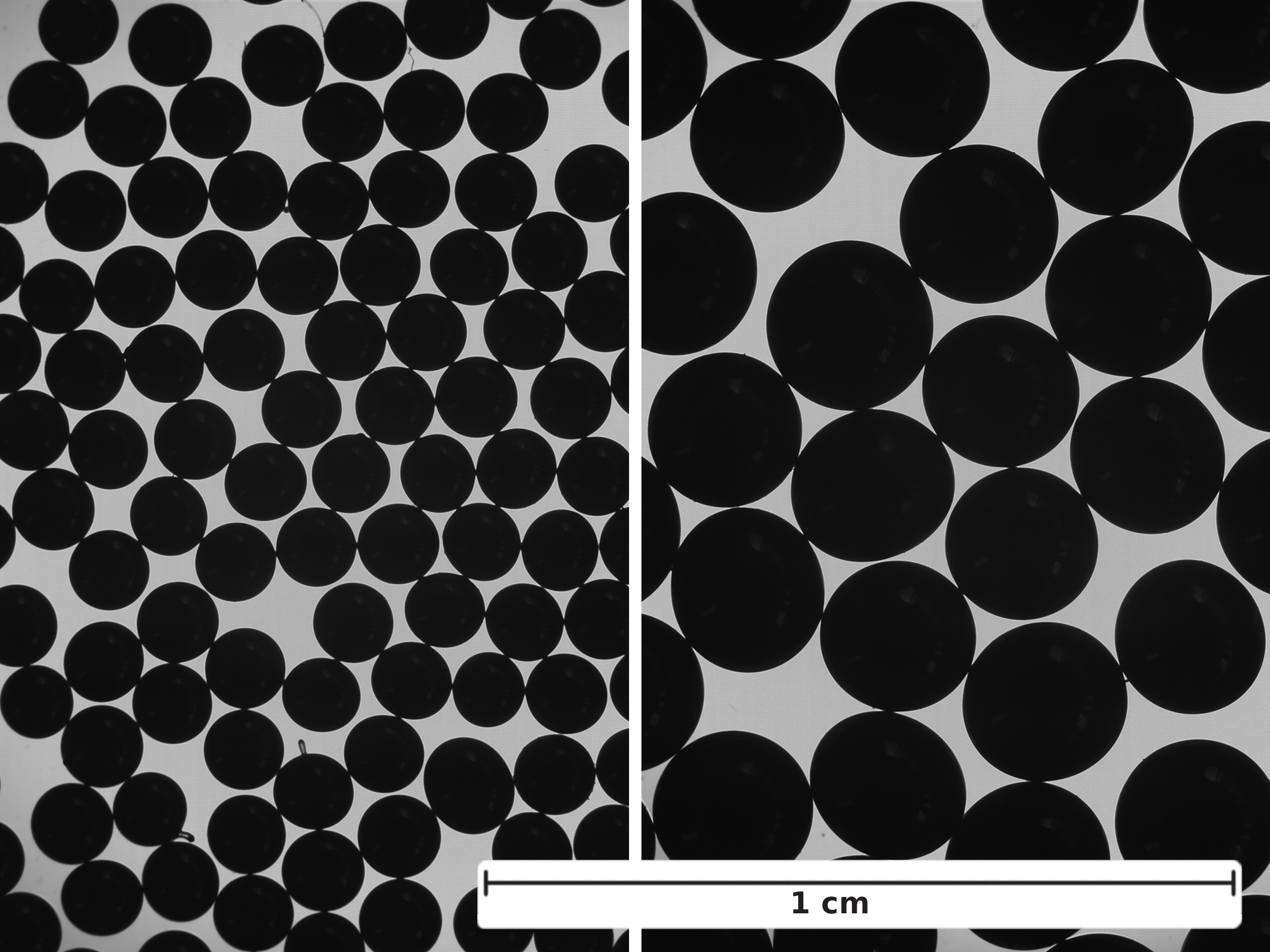}
	\caption{\label{fig.foto}Bright field image of the basalt grains used visualizing the monodispersity and sphericity.}
\end{figure}
We used these samples as they could be acquired easily (purchases from Whitehouse Scientific), as collisions provide enough charges for easy measurements of individual grains, and as the sample is rather monodisperse. Especially the latter allows us to concentrate on pressure effects without considering biased charging due to size differences.}

\begin{figure}[h]
	\includegraphics[width=\columnwidth]{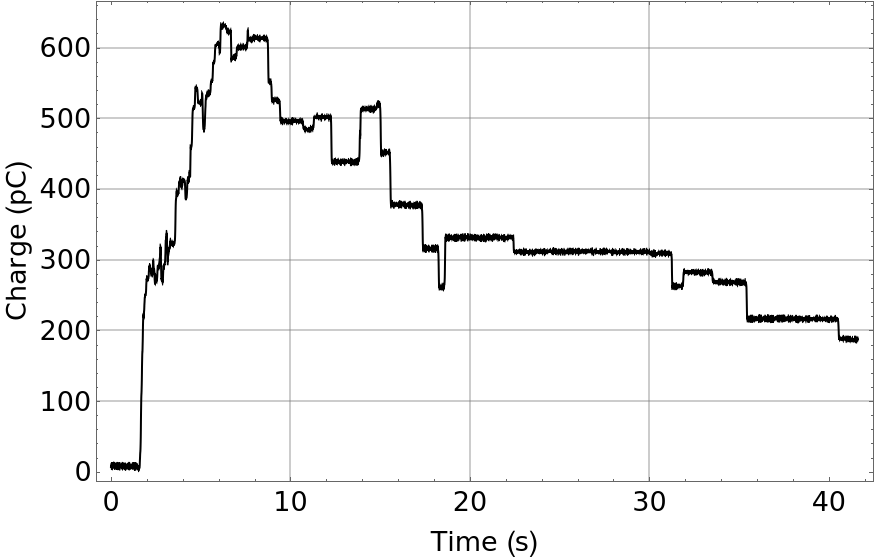}
	\caption{\label{fig.chargemeasurement}Example of a charge measurement of all grains extracted from a sample collisionally charged before. {Every step marks one particle. The particles leave in faster sequence initially, while the container is still filled with many grains. All grain charges are still resolved individually by the electrometer. The tendency that positive charges leave first in this example is not systematic but occurs by chance.} }
\end{figure}

 About 50 particles were used for each experiment. At the center of the sample cylinder a hole of 1.6 mm or 2.8 mm diameter, depending on the sample, allows individual grains to leave the shaker if a lid is removed.
Passage of a grain is signaled by a light barrier. 
Only one particle at a time is allowed to move out. 
{Contact of the grain with a non-particle surface during exit is possible but we expect only a minor charge transfer as only few contacts can occur, covering a relatively small surface area in contrast to the charging process during vibration of the spheres before.}
Grains then drop into a Faraday cup which is attached to an electrometer (Keithley B2985A). This allows a measurement of the charge of each grain. The shaker and Faraday cup are placed into a vacuum chamber to allow a pressure dependent charge measurement. 
An example of a charge measurement is shown in fig. \ref{fig.chargemeasurement}.
Charge differences at the steps mark individual grains with their respective charge.
{In this example, the first grain is strongly positive, essentially carrying an excess charge after all spheres left the container. This can happen as the grains fixed on the inside of the container also carry a charge. This does not change the picture of the charge distribution though which is centered around zero as seen below.}

\section{Results}

\begin{figure}[h]
	\includegraphics[width=\columnwidth]{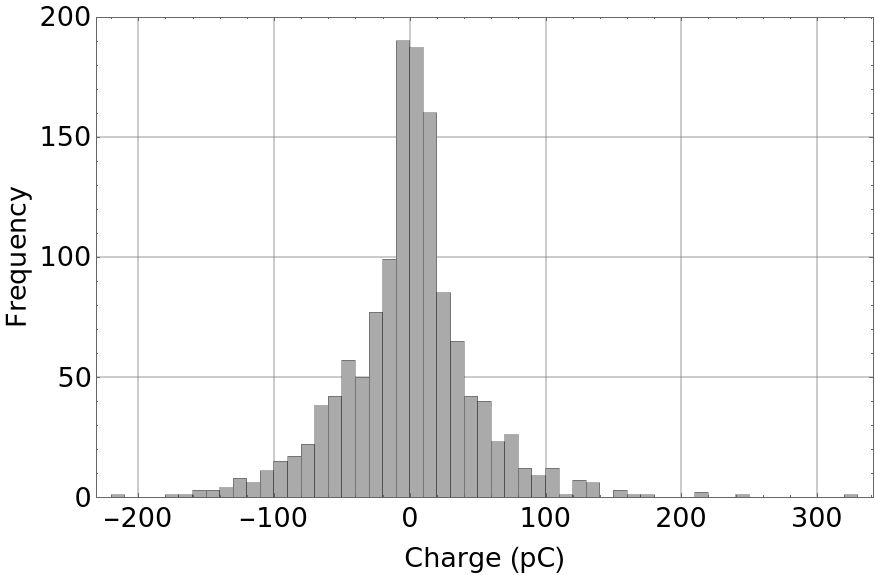}
	\caption{\label{fig.distribution}Charge distribution of all experiments from the 2 mm sample. {In total, charges from 22 experiments at all pressures are summed to get a smooth curve. This shows that overall the charge distribution is symmetric and essentially has zero net charge.} }
\end{figure}

For each measurement at a given pressure we get a charge distribution. The distribution of all charges measured is shown in fig. \ref{fig.distribution}.
The charge distribution is essentially symmetric around 0 charge with only a negligible offset. The overall neutrality is is consistent with same kind of particles randomly interacting with each other. {\citet{Harper2017}, \citet{Harper2016} or \citet{Forward2009} show in similar experiments that a few minutes are sufficient to reach charge equilibrium.} We therefore consider the shaking time long enough that the charge distributions are equilibrium distributions.

\begin{figure}[h]
	\includegraphics[width=\columnwidth]{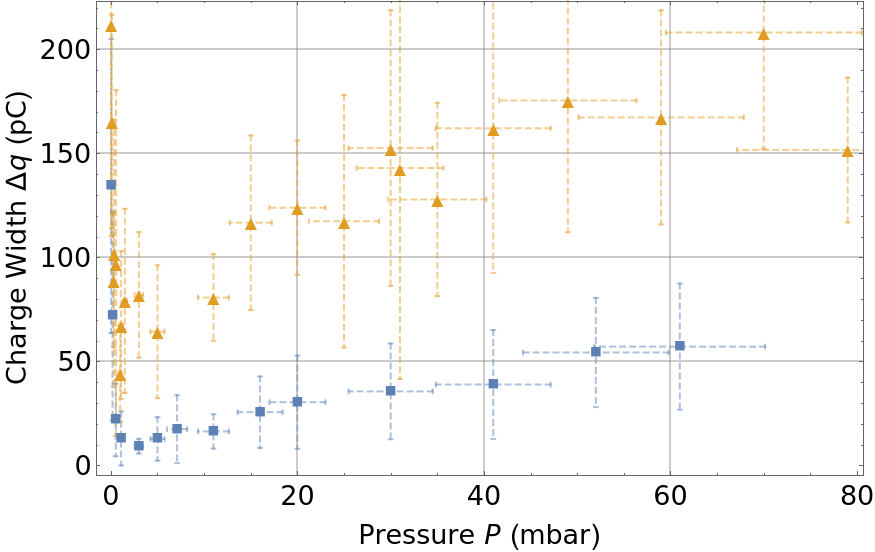}
	\caption{\label{fig.chargemeasurement7}Maximum charge depending on ambient pressure; blue squares: grains of 1 to 1.2 mm in size; orange triangles: grains of 2 to 2.4 mm in size; Error bars in x are uncertainties of the pressure measurement (sensor); Error bars in y are standard deviations.}
\end{figure}

As pressure dependent quantity we consider {the average charge and standard deviation of the upper and lower fourth of the measured charges} for a given measurement. This avoids weighing individual outliers too much. For distributions centered at zero, essentially the case here, this standard deviation then essentially equals the absolutes of the maximum charges, which we will call $\Delta q$. 
The data on maximum charging are seen in fig. \ref{fig.chargemeasurement7} for all pressures measured.
It is clearly visible that there is strong dependence of the maximum charges achievable on pressure, which at first glance shows the same trend as a typical Paschen discharge curve which we will model in the following section.

\section{Discharge Model}

Here, we set up a model that describes the discharge in agreement to the experimental data.
Basic idea is that two grains with different charge can discharge if they rebound from or approach  each other if conditions for discharge are given. This essentially means, that charge is building up in many collisions until such conditions are met. 
Under this premises, the discharge condition can be quantified by 1) using the Paschen law based on the Townsend breakdown mechanism and 2) considering the Coulomb potential of two charged spheres with distance.

The Paschen law describes the breakthrough voltage $U_b$ applied between two plate capacitor electrodes depending on distance between the electrodes $d$ and the pressure $P$

\begin{equation}
U_b = \frac{B \cdot Pd}{ln(A \cdot Pd) - ln \left( ln \left( 1+\frac{1}{\gamma} \right) \right)}
\label{break}
\end{equation}

Here, $A$ and $B$ are constants depending on the gas used, 
$\gamma$ is the probability that an electron produces a new ion upon impact on the electrode. As we consider the discharge between spheres, we assume that the same type of equation holds if the electrodes were not plates but spherical particles being charged homogeneously.
For the simplest model we further assume that the voltage between separating, spherical grains $U$ is a Coulomb potential of two homogeneously charged particles.

\begin{equation}
U =  \frac{2 \Delta q}{4 \pi \epsilon_0} \left( \frac{1}{r_0} - \frac{1}{r_0 + d} \right)
\label{coulomb}
\end{equation}

\begin{figure}[h]
	\includegraphics[width=\columnwidth]{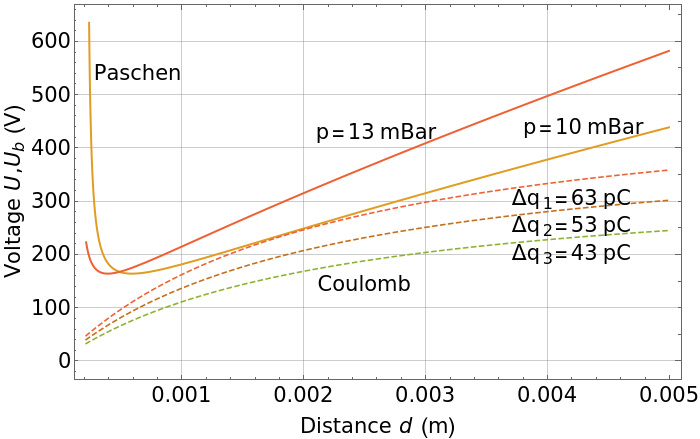}
	\caption{\label{discharge.model.1}Main model idea: {Two breakdown volatages at 10 mbar and 13 mbar are shown (eq. \ref{break}); Constants used are for air (A = 10.95 $\rm (Pa \cdot m)^{-1}$, B = 273.8 $V / (Pa \cdot m)$, $\gamma = 0.1$); Radius is $r_0 = 2.2$ mm; In addition three Coulomb potentials are shown for three charges (eq. \ref{coulomb}); The maximum charge allowed on a grain corresponds to the case that Coulomb voltage and breakdown voltage touch. In this case 63 pC is allowed at 10 mbar; For 13 mbar higher charges would be possible.}}
\end{figure}

Here, $r_0$ is the grain radius.
Examples of $U$ and $U_b$ are shown in fig. \ref{discharge.model.1}.
If both voltages equal each other as the grains' charges increase, breakdown and partial discharge occurs. This sets the maximum charge $\Delta q$. Equaling eq. \ref{break} and \ref{coulomb} this charge can be calculated depending on $P$, $d$, and $\gamma$.

\begin{equation}
\Delta q  = \frac{2 \pi \epsilon_0 B Pd}{ln(APd)-ln(ln(1+1/\gamma))}\left( \frac{1}{r_0}-\frac{1}{r_0+d}\right)^{-1}
\end{equation}

The constants $A=10.95 \rm (Pa \cdot m )^{-1}$  and $B = 273.8 \rm V/(Pa \cdot m )$ are taken for air \citep{Lehr2017}. For a given pressure $P$ and given $\gamma$ the {smallest charge $\Delta q$} for which this equation can be fulfilled gives the maximum charge {that can build up} before a discharge can occur.
For higher charges both functions do not touch but cross and smaller charges would already lead to breakthrough, so the tangential point is the important one, the final result will be the same though. We note this here as e.g., \citet{Harper2016} consider a very similar setting but start with higher charges after separation, crossing the Paschen curve. Eventually, the conditions at the tangential point are the stable ones. We also note though that this simple picture should not be overstressed as in any case the charge distribution will not be homogeneous on the surface and also the orientation and charge distribution of the grains will be important in detail. 

\begin{figure}[h]
	\includegraphics[width=\columnwidth]{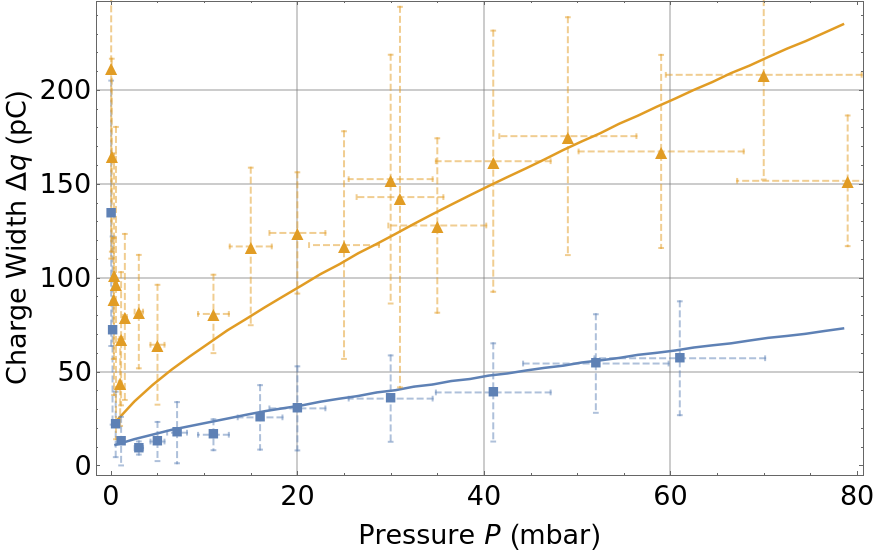}
	\caption{\label{fig.badmodel}Data and simple model for the maximum charges achievable at a given pressure based on Paschen curve and pure Coulomb potential; blue squares: grains of 1 to 1.2 mm in size; orange triangles: grains of 2 to 2.4 mm in size; Error bars in x are uncertainties of the pressure sensor; Error bars in y are standard deviations. {The model lines are no fits but calculated using the procedure and typical parameters given in fig. \ref{discharge.model.1} which just match the data.}}
\end{figure}

In any case, {using this procedure, we can determine one $\Delta q$ for every pressure. If we use $\gamma=0.1$ as typical value there is no free parameter left} to describe the high pressure branch of the experimental data.  
The resulting pressure dependence of the maximum charge is shown in fig. \ref{fig.badmodel}.
Though the model is rather simple, it is in good agreement to the high pressure branch of the experiments which suggests that the idea captures the main mechanism.
However, this cannot explain the low pressure dependence of the data.
This is intrinsic to the assumptions and functional dependences of the Paschen and Coulomb laws.
 
We therefore modify the model and introduce a screening of the individual charges for larger distances. We therefore assume that the electrical field decreases with distance, as the electrical charge of one sphere is screened by other (charged) particles in the vicinity. This is reasonable in a larger sample of grains as the total sample of identical grains is neutral and the field at larger distances is the average of many fields while shortly after separation the colliding grains only see each other's field. 
{Similar to a general plasma shielding} we assume a simple exponential distance dependence and use a modified effective Coulomb voltage. 

\begin{equation}
U_{mod} = U \cdot e^{-\frac{d}{\lambda}}
\label{modifiedmodel}
\end{equation}

\begin{figure}[h]
	\includegraphics[width=\columnwidth]{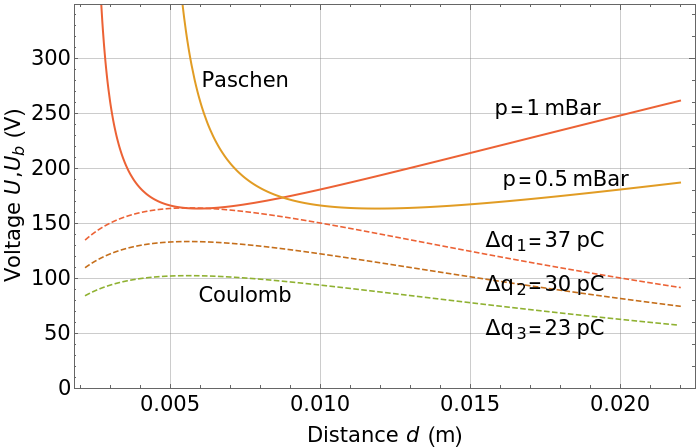}
	\caption{\label{fig.chargemeasurement9}Modified model: Paschen curve and voltage with exponential decay for 3 different charges. {Same parameters as used in fig. \ref{discharge.model.1} and a screening constant $\lambda = 2$ cm (eq. \ref{modifiedmodel}); Again, the maximum charge allowed is determined by the case if both curves touch. Here, at 1 mbar this is $36.98$ pC}}
\end{figure}

\begin{figure}[h]
	\includegraphics[width=\columnwidth]{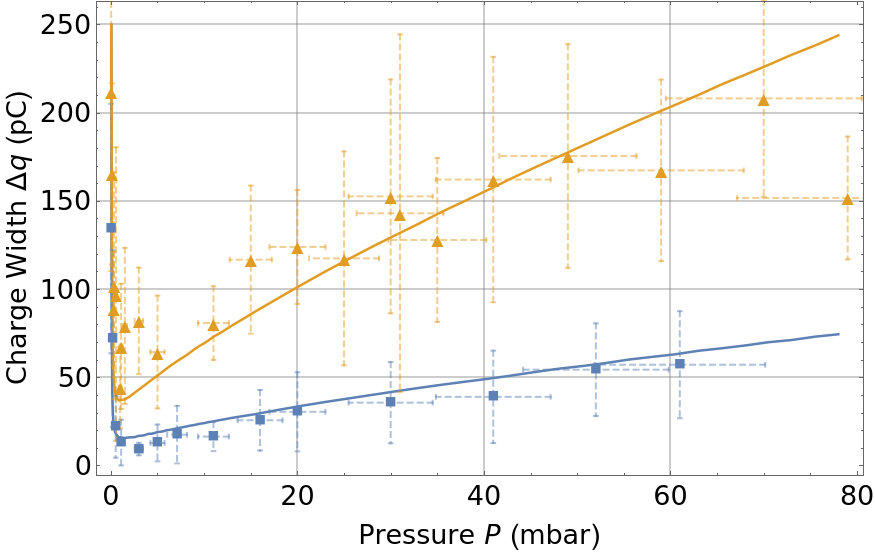}
	\caption{\label{fig.chargemeasurement2}Experimental data and model with exponentially decaying potential; blue squares: grains of 1 to 1.2 mm in size; orange triangles: grains of 2 to 2.4 mm in size; Error bars in x are uncertainties of the pressure sensor; Error bars in y are standard deviations; {The model line is calculated using the parameters and procedure visualized in fig. \ref{fig.chargemeasurement9}}}
\end{figure}

The parameter $\lambda$ gives the characteristic screening length {comparable to a Debye-length}. A value that works well
is $\lambda = 2\,\mathrm{cm}$, which again is reasonable with the given number of particles and size of the container. We use this value without further detailed fitting.
Again finding the lowest charge (numerical) for which the breakthrough voltage curve is touched we find the situation as visualized in fig. \ref{fig.chargemeasurement9}.
As the potential now decreases towards larger distances also conditions at low pressure can be found where  breakthrough occurs at higher charge values for small distances. Calculating again the tangential points and respective charges for all pressures studied, we find fig. \ref{fig.chargemeasurement2}.

We note, that with an electrical field not changing sign with distance, even a screened Coulomb based voltage can actually not really decrease with distance. On the other side, also the breakthrough voltage will change its distance behavior in a screened case. So this exponential decay with distance is currently only a simple mathematical description that fits the data well with breakthrough as main mechanism to limit charging.

\section{Discussion}

We clearly see a difference in charging depending on pressure. The maximum charge on a grain is smallest for the pressure range relevant for the Martian surface. In terms of charge density we find the lowest values to be $7 \cdot 10^{-6}\rm C / m^2$ (using $\Delta q = 10$ pC) for the 1 mm grains and $11 \cdot 10^{-6}\rm C / m^2$ (using $\Delta q = 60$ pC) for the 2 mm grains which is on the same order. The highest values measured can be a factor of 10 higher or about  $10^{-4}\rm C / m^2$.

These values might be compared to maximum charge densities found in other experimental works, keeping in mind though that experimental conditions are never the same. 
E.g., for 2 silica surfaces with one coated with amino-silane \citet{Horn1993} find  $1.7 \cdot 10^{-3}\rm C / m^2$ after discharges in dry nitrogen atmosphere, which is about a factor 10 higher than the highest values we find. Some of this might be attributed to their higher pressure of 1 bar as our data and the model are still rising significantly at 80 mbar. The remaining small factor might be due to geometry and material differences. \citet{Poppe2000} deduced a similar value of about $10^{-4} \rm C / m^2$  in a microparticle contact with a flat silica surface at a pressure of about 0.01 mbar. These values are in agreement to our data by an order of magnitude. At the same time they give a much lower value for their silica target of only 9 $\rm 10^{-6} C / m^2$ but consider this to be due to charge spreading in the target.
As last example we mention the recent work by \citet{Haeberle2018}. They used quite a number of materials for particle-wall charging experiments at normal pressure.
Like \citet{Poppe2000}, based on Hertzian contacts, they deduce an average of 100 $\rm e^- / \mu m^2$ or about $1.6 \cdot 10^{-5}\rm C / m^2$. The highest values also reach about $ 10^{-4}\rm C / m^2$ though, again with some variations due to materials. Overall, the range of maximum charge densities that can be reached before discharge occurs are consistent but can vary by a factor 10 or more, e.g. due to pressure variations. 

We note that our model assumes the charge to be homogeneously distributed on a sphere. 
The screening is currently still more of a phenomenological approach. This might depend on the particle density and the surface of the shaker. It might not be present in a very dilute granular gas but the current experiment cannot answer that. For application to Mars the screening mechanism is not of relevance as it only describes the pressure region below Martian values. 
{Screening, contact sizes and other properties are depending on grain size so the details of charging and discharge might change with grain size if changed by orders of magnitude, e.g. considering small micrometer dust.}

In application to Mars, there might be a competition between lower local grain charging and easier large scale discharge in the thin atmosphere.  However, also the total number density of grains that the thin atmosphere can sustain and the increased sedimentation, favoring separation but also  loss of grains might be important. {\citet{Farrell1999a} considered that a corona-like discharge of a dust grain might occur in the Martian atmosphere, limiting the charge on an individual dust grain. In any case, \citet{Farrell1999a} simulating charge separation and discharge in the Martian atmosphere find that some filamentary discharge should still be possible. \citet{Jackson2006} measured large electric fields in dust devils on Earth and speculate on a discharge in Martian dust devils as potential dissipative glow discharges, if the same contact electrification occurs. We study the contact electrification as one of the important factors in detail here.
\citet{Forward2009b} already studied triboelectric charging of grains with Mars in mind, i.e. using Martian dust analog (JSC-1 Mars). They showed that the dust grains, being between about 100 $\rm \mu m$ and 1 mm in size, charged according to grain size. The small grains charged negative, the large grains charged positive in agreement to a trapped electron state model.} 

{In contrast, we consider a monodisperse, homogeneous sample here to avoid any bias due to size difference or chemical composition to see the pure effect of ambient pressure on grain charging.
Observations of discharges up to a cm scale at Martian pressures was already observed in laboratory settings by \citet{Mills1977}. More recently, \citet{Krauss2003} also did experiments and varied the ambient pressure in triboelectric experiments in the range of Martian pressures and observed discharge in a sheared granular medium using JSC-1 Mars simulant. In addition they observe discharges when dropping pre-charged material of a different density onto the regolith. They did observe a pressure dependence with the number of discharges decreasing with pressure, i.e. a high rate at 1 Torr and much smaller rates from 2 to 7 Torr. 
The setting is different from ours, including different materials, different sizes and a dense environment during charging. In view of our results we  interpret their findings as being based on a charge reduction on individual grains at mbar pressures with additional effects due to the nature of the experiments.}

As result, small scale discharges between two grains certainly take energy out of the electrical system. So whatever follows, even if any discharge on larger scales occurs later, the energy content would strongly be decreased. The data show a difference of about a factor of 5 between maximum charges achievable at a few mbar and 80 mbar.
We restricted the measurements to this low pressure region as the absolute water content is low in that case. At higher pressures measurements are sensitive to humidity. Certainly, the charging still increases to larger pressures and conservatively estimated, the difference might rather have a lower limit of a factor of 10 for charging comparing Mars and Earth. 
{As different story, the results might also be of importance in chemical processes occurring on surfaces of Martian dust grains due to triboelectric effects, where breakdown is also an important parameter \citep{Tennakone2016}.}

\section{Conclusion}

The experiments show a clear limit of the maximum collisional charging of grains depending on the ambient pressure. We used basalt, being a typical mineral on Earth as well as on Mars. We find that grains charge to a level that is a factor of 5 smaller under Martian conditions compared to 80 mbar and likely a factor of 10 smaller than on Earth. This can well be explained by small scale discharges upon rebound of charged grains. Without considering all aspects of lightning generation {or large scale discharges} this does not rule out lightning. Grains still get charged. However, this limits the capabilities for generating  lightning and related phenomena strongly and in case of such a discharge the energy released will be an order of magnitude lower.
The small scale discharges preventing high grain charges might well be one reason why it is still debated if lightning on Mars occurs at all.\\

\section{acknowledgements}
This project is supported by DLR Space Administration with funds provided by the Federal Ministry for Economic Affairs and Energy (BMWi) under grant number DLR 50 WM 1762. We appreciate the reviews by the two anonymous referees which strongly helped improving this manuscript.

\bibliography{bib}

\end{document}